\documentclass[reprint,
nofootinbib,
 amsmath,amssymb,
 aps,
]{revtex4-2}

\usepackage{graphicx}
\usepackage{dcolumn}
\usepackage{bm}
\usepackage{braket}
\usepackage{xcolor}
\usepackage{footnote}
\usepackage[normalem]{ulem}
\usepackage{mathtools}
\newcommand\underrel[3][]{\mathrel{\mathop{#3}\limits_{%
      \ifx c#1\relax\mathclap{#2}\else#2\fi}}}

\begin{document}

\title{Quantum Information Engines: Assessing Time, Cost and Performance Criteria}

\author{Henning Kirchberg}
\email{khenning@sas.upenn.edu}
\author{Abraham Nitzan}
\affiliation{
 University of Pennsylvania, Department of Chemistry, Philadelphia, PA, 19104
}

\date{\today}

\begin{abstract}
In this study, we investigate the crucial role of measurement time ($t_m$), information gain and energy consumption in information engines (IEs) utilizing a von-Neumann measurement model. These important measurement parameters allow us to analyze the efficiency and power output of these devices. As the measurement time increases, the information gain and subsequently the extracted work also increase. However, there is a corresponding increase in the energetic cost. The efficiency of converting information into free energy diminishes as $t_m$ approaches both 0 and infinity, peaking at intermediate values of $t_m$. The power output (work extracted per times) also reaches a maximum at specific operational time regimes. By considering the product of efficiency and power as a performance metric, we can identify the optimal operating conditions for the IE.
\end{abstract}

\maketitle

\textit{Introduction-} Heat engines operate between reservoirs at different temperatures, extracting useful work while unavoidably dissipating non-useful heat. Alternatively, a single heat bath may be used as the energy source in feedback controlled devices \cite{sag2008,sag2010,sag2012,vid2016,cot2017,Cottet2018,elo2017,elo2018,mon2020,bre2021,ZurBook,mar2009,man2012,def2013,str2013,hor2013,bar2013}, referred to below as information engines (IEs), in which information about the system's state is obtained by some "Maxwell demon" and used to control the engine's operation \cite{MaxwellBook,ZurBook}. The second law is accounted for by the entropy increase during the demon's restoration to its initial state, also implying a minimal added operation cost, Landauer's erasure work \cite{lan1961}. In the quantum version of such devices the demon's acquisition of information is often described as a quantum measurement process with a prescribed action on the system, often utilizing positive operator-valued measures (POVM) with Kraus operators while disregarding the actual physical nature [See, e.g. \cite{ann2022,och2018,wis1994}]. Such an approach makes it possible to investigate important thermodynamics characteristics of information engines (such as the aforementioned Landauer lower bound on the unavoidable dissipation, which recent studies have shown to be compatible with fluctuation theorems of stochastic thermodynamics \cite{sag2010,toy2010,sag2012}). However, these measurement descriptions cannot be easily used to study important performance characteristics such as efficiency and operation power because standard measurement theory rarely considers the energy cost of measurement in relation to the measurement time (see Refs. \cite{ghi1979,bra2013,bus2010,bre2021} regarding the latter).

This letter addresses these important issues for the first time by (a) describing the information acquisition process utilizing a Von-Neumann quantum measurement model \cite{VonNeumannBook} in which the meter is part of the system, making the measurement time an achievable system parameters and making it possible to calculate the measurement energetic cost; (b) using these time and cost observables to define and calculate the efficiency and operating power of the IE; and (c) comparing the performance of our IE model to that of a standard heat engine and identifying conditions under which the IE is advantageous. While these issues are studied within a particular IE model, our approach is applicable to a wide range of such models and highlights the need to address information acquisition as a time-dependent physical process when assessing their performance.

\begin{figure}[h!!!!!]
\centering
\includegraphics[width=\linewidth]{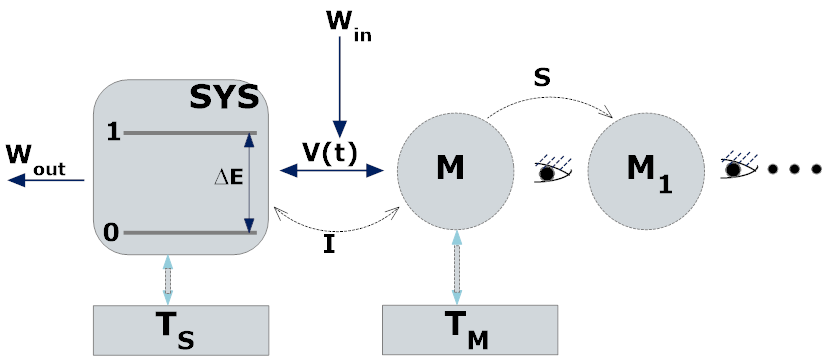}
\caption{\label{fig1A} Schematics of an IE model. A system (SYS) and a meter (M), each coupled to their own thermal bath of temperatures $T_{S}$ and $T_M$, respectively, are entangled by an interaction $V(t)$. The state of M is projectively monitored by another meter M$_1$ accompanied by an entropy flow $S$ between M and M$_1$. The information $I$ is used to extract energy $W_{out}$ form the system bath. The measurement time and its energy cost $W_{in}$ are computed and used to calculate the energy efficiency and operating power.}
\end{figure}
 
\textit{IE model-} In our IE model (Fig.\ \ref{fig1A}) the working entity (SYS) is a 2-state system (2SS) with the energy of its lower eigenstate $\ket{0}$ set as zero and the upper state $\ket{1}$ at energy $\Delta E$. This system is monitored by coupling it (the interaction $V(t)$ in Fig.\ \ref{fig1A}) to a meter M modeled as an otherwise free particle. The Hamiltonian of this combined 2SS-M-system reads $\hat{H}=\Delta E \ket{1}\bra{1}  +\frac{\hat{p}^2}{2} + \hat {V} (t)$ where $\hat{p}$ is the mass weighted momentum operator. In the present analysis we take $\hat{V}(t)\equiv \hat{V}=\hat{x}\otimes\ket{1}\bra{1}$, for $0\leq t \leq t_m$, and $\hat{V}(t)=0$ otherwise, so that during the measurement time $t_m$ the meter responds to the system only if the latter is in state 1. This on/off switching requires work by an external agent which is part of the measurement energetic cost. In addition, the system and meter are coupled to their thermal environments of temperature $T_S$ and $T_M$ respectively, although in the present analysis we take, following \cite{jor2020}, $T_M=0$, and set the meter initial state to be a free particle wavepacket, Eq.\ \eqref{meterIn} below. Further, we assume that the momentum state of the meter can be instantaneously and projectively determined by a Maxwell demon (M$_1$) and, because, $T_M=0$, we can disregard the Landauer overhead cost ($T_M S$) for entropy increase $S$ associated with erasure of the demon's memory related to this step. 

Finally, based on the information obtained by monitoring the meter's state, energy is extracted from the 2SS. A way to do it is the stimulated emission concept suggested in Refs.\ \cite{bre2021,elo2017}: if the 2SS is known to be in its upper state, a resonant $\pi/2$ pulse can bring it to the ground state while extracting a photon. The average (over engine cycles) excess energy that can be extracted using such information from the 2SS is denoted $W_{out}$ in Fig.\ \ref{fig1A}. On the cost side, in addition to the energy associated with controlling $V(t)$ we have to account for the energy needed to prepare the initial meter state. The detailed IE engine cycle is described by the following steps: 

(i) \textit{Initial state:} The 2SS is taken to be initially in thermal equilibrium with a bath at temperature $T_S$. The meter is taken to start at a wavepacket state of the form  
\begin{align}
\label{meterIn}
D(p)=\braket{p|D}&=\bigg( \frac{2}{\pi \hbar^2 B} \bigg)^{1/4} e^{-p^2/\hbar^2 B} 
\end{align} in the momentum representation, whose width $B$ represents a lower bound on its inverse size (in mass weighted length units). This defines the initial density matrix of the combined system as 
\begin{align}
&\hat{\rho}(t=0)=\hat{\rho}_S(t=0)\otimes \hat{\rho}_M(t=0),
\end{align}
with
\begin{align}
\label{in1} 
\hat{\rho}_S(t=0)= a\ket{0}\bra{0}+b\ket{1}\bra{1}; \hspace*{0.1cm} \hat{\rho}_M(t=0)= \ket{D}\bra{D}, 
\end{align}
where $a$ and $b$ are real positive numbers satisfying $T_S=\Delta E/k_B [\ln(a/b)]^{-1}$ and $a+b=1$.

(ii) \textit{Entangling evolution:}  During the time interval $(0,t_m)$, the system and meter evolve under the Hamiltonian $\hat H$, leading to an entangled state described by the density matrix $\hat{\rho}(t_m)=e^{-i \hat{H}t_m/\hbar} \hat{\rho}(0) e^{i \hat{H}t_m/\hbar}$.

(iii) \textit{Projective measurement \& information gain:}  Following the entangling evolution, the state of the meter is projectively determined to be the eigenstate $\ket{p}$ of the momentum operator. It is assumed that this process is instantaneous and involves no energy cost \cite{footnote5}. The conditional probability of the 2SS to be in state $i=0;1$ given the meter outcome $p$ is thus determined by $P_i(t>t_m|p)=\bra{i} \hat{P}(p,t_m)\ket{i}/{Q(p,t_m)}$ where $\hat{P}(p,t_m)\equiv \braket{p|\hat{\rho(t_m)}|p}$ and $Q(p,t_m)=\sum_{i=0}^1 \bra{i} \hat{P}(p,t_m) \ket{i}$.  $\hat{P}(p,t_m)$ is the  joint system-meter density operator to be in state $i=0;1$ and measuring the meter outcome $p$, which reads
\begin{align}
    \hat{P}(p,t>t_m)=P_0(p,t) \ket{0}\bra{0}+ P_1(p,t) \ket{1}\bra{1},
\end{align}
where for our IE model,
$P_0(p, t)= \sqrt{\frac{2}{\pi \hbar^2 B}}   a e^{-\frac{2p^2}{\hbar ^2 B}}$ and $P_1(p, t)= \sqrt{\frac{2}{\pi \hbar^2 B}} b e^{-\frac{2(p+g t)^2}{\hbar ^2 B}} $ \cite{SM1}. 
 
\begin{figure}[h!!!!!]
\centering
\includegraphics[width=\linewidth]{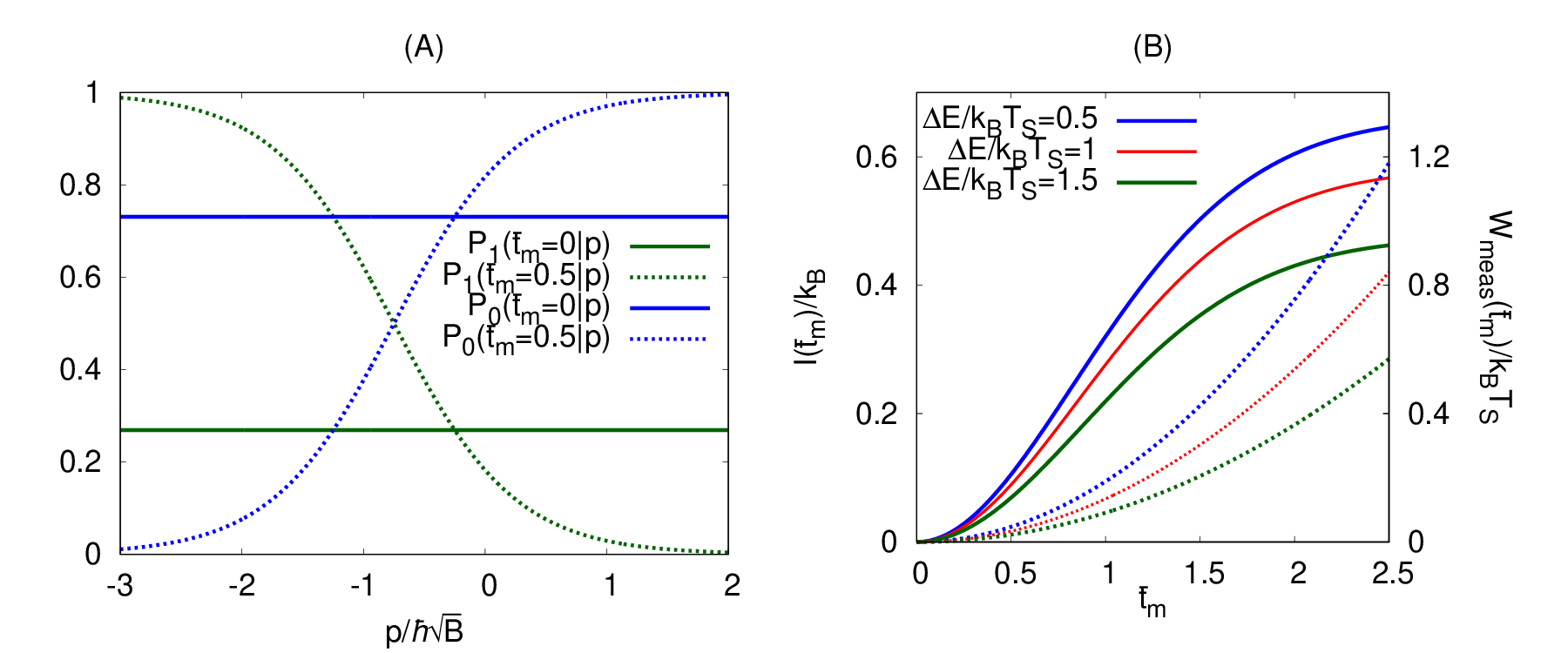}
\caption{\label{fig9} (A) The conditional probability $P_{i=0;1}(\bar{t}_m|p)$ that a 2SS is in state $0$ or $1$ given that the meter outcome is $p$. The 2SS is initially in thermal equilibrium with $T_S$. The parameters used are $T_S=300$K, $\Delta E=k_B T_S$ and the initial meter state is given by Eq.\ \eqref{meterIn} with $\hbar^2 B=25.85$meV (This choice of B corresponds to $W_{prep}=k_BT/4$ with $T=300$K). The horizontal lines represent the system's initial state, and the dotted lines are for $\bar{t}_m=0.5$. (B) The information gain $I(\bar{t}_m)$ (solid lines, left axis) and the measurement energy cost $W_{meas} (\bar{t}_m) $ (dotted lines, right axis) plotted against measurement time $\bar{t}_m$ for different choices of $\Delta E/k_B T_S$ with $\hbar^2B=25.85$meV and $\Delta E=25.85$meV (which corresponds to $\Delta E=k_BT$ with $T=300$K).}
\end{figure}

The information gain, $I(t_m)$, in this measurement process can be quantified by averaging the conditional system entropy
$S(t_m|p)=-k_B \sum_{i=0}^{1}  P_i(t_m|p) \ln{P_i(t_m|p)}$ over an ensemble of identical measurements, $S(t_m) = \int dp Q(p,t_m) S(t_m|p)$, leading to \cite{footnote4} 
\begin{align}
\label{know}
I(t_m)\equiv S(0)-S(t_m),
\end{align}
where $S(t_m)= -k_B \int dp \sum_{i=0}^{1}  P_i(p,t_m) \ln{P_i(t_m|p)}$ and $S(0)=-k_B(a \ln a + b \ln b)$. $S(t_m)$ is a monotonously decreasing function of the measurement time that vanishes at $t_m\to \infty$, showing that the information gain monotonously increases with $t_m$ towards its maximum $S(0)$ in the standard Shannon interpretation \cite{sha1948} (see Fig.\ \ref{fig9} (B) and later discussion). 

In the analysis below we assume that the measurement time $t_m$ dominates the IE cycle time, namely that the time involved in other processes is relatively negligible as will be discussed later. The magnitude of $t_m$ might be determined by the spatial range of the system-meter interaction in realistic set-ups.

(iv) \textit{Work extraction:} We assume, in an idealized setup, that if the 2SS is known to be in its excited state, this excitation energy can be fully extracted (eg., by stimulated emission \cite{elo2017,mon2020,bre2021}). Additionally, any attempt to extract this energy involves unknown costs. The need to consider such costs may be circumvented by focusing on the excess energy gain, which for the meter outcome $p$ is given by \cite{footnote13}
\begin{align}
\label{gainEvent}
  G(p,t_m)=\Delta E [P_1(t_m|p)-P_1(0|p)] =   \Delta E [P_1(t_m|p)-b].
\end{align}
Depending on the meter outcome $p$, $G(p,t_m)$ can be negative, while the average over all possible meter outcomes vanishes
\begin{align}
\label{gain}
W_{out}(t_m) &= \int_{-\infty}^{\infty} dp Q(p,t_m) G(p,t_m) =0.
\end{align}
To see this, note that the integrand in Eq.\ \eqref{gain} is the difference between the joint probabilities $P_1(p,t_m)-P_1(p,0)$ and the integral over all $p$ just yields the constant probability that the 2SS is in state $i=1$ irrespective of the meter outcome. A productive use of the information engine is achieved by restricting the photon extraction attempts to events for which the measurement outcome $p$ indicates that the 2SS probability to be in the excited state 1 is large enough relative to its thermal value. For our model these are events in which the meter outcome is smaller than some bound $p<p'$ where $-\infty < p' < 0$ (see Fig.\ \ref{fig9} (A) and discussion below). In this case the averaged useful energy extracted per attempt is $\bar{W}_{out}(t_m,p')=\int_{-\infty}^{p'}dp \bar{Q}(p,t_m)G(p,t_m)>0$ where $\bar{Q}(p,t_m)=Q(p,t_m)(\int_{-\infty}^{p'} dp Q(p,t_m))^{-1}$. However, not every engine cycle ends with an extraction attempt. We may define the average effective cycle time by $ t_{eff} = ( \int_{-\infty}^{p'} dp Q(p,t_m))^{-1} t_m$ \cite{footnote14}.
The average useful energy extracted per cycle is therefore
\begin{align}
\label{gain2}
W_{out}(t_m,p') &= \bar{W}_{out}(t_m,p') (t_m/ t_{eff})  \\ \notag &= \int_{-\infty}^{p'} dp Q(p,t_m) G(p,t_m) >0.
\end{align}

(v) \textit{Restoration:} Following the measurement-informed extraction of useful energy, the engine cycle is closed by restoring the 2SS and meter to their initial states. The former is brought back to equilibrium with its thermal environment at temperature $T_S$. Because the photon extraction step leaves the 2SS in its lower (ground) state $\ket{0}$, restoring thermal equilibration involves heat transfer from the bath which on the the average must be equal to the gain \eqref{gain2}.  For the meter, a route for preparing it in the initial state $\ket{D}\bra{D}$ (Eq.\ \eqref{meterIn}) may start by bringing it to equilibrium with a zero temperature bath, followed by adiabatically confining it in a harmonic potential for which Eq.\ \eqref{meterIn} is the ground state and then suddenly releasing the confinement leaving a free particle in the state $\ket{D}\bra{D}$. Such a route requires an energy investment equal to the zero point energy (denoted $W_{prep}$ in Eq.\ \eqref{invest2}) of a particle in the harmonic confining potential. This value is used below as the energy cost of the meter restoration step.

Consider next the energy cost of the cycle described above which is the sum of two contributions: (a) the energy needed to create the meter system entanglement and (b) the energy associated with the restoration of the meter to its initial state \cite{footnote2,footnote12}. 

(a) The energy needed to create the system-meter entanglement is henceforth referred to as the measurement energy, $W_{meas}$. Because energy is conserved during the unitary evolution of the interacting 2SS and meter, the cost of this process is associated with  switching the system-meter interaction on and off, which is given by
\begin{align}
\label{intrE}
W_{meas}(t_m) \equiv \textrm{tr} [\hat{\rho}(0) \hat{V}]-\textrm{tr}  [\hat{\rho}(t_m) \hat{V}] ,
\end{align}
where $\textrm{tr} [\dots]\equiv \int dp \sum_{i=0,1} \bra{p} \bra{i} \dots \ket{i}\ket{p}$. This energy may be thought of as the work done by the agent who switches the interaction on and off (more generally, who affects the time-dependence of $\hat{V}$). In writing Eq.\ \eqref{intrE} we have assumed that this switching is instantaneous. For our choice of initial states and system-meter interaction $\textrm{tr} [\hat{\rho}(0) \hat{V}]=0$, namely switching on the interaction costs no energy. 

(b) Secondly, some of the energy difference between the meter in its post-measurement state (a free particle with momentum $p$) and the meter initial state, Eq.\ \eqref{meterIn}, could be extracted as useful work. This work can be subtracted from the energy cost in \eqref{intrE}. Alternatively, following Landauer \cite{lan1961}, we may regard the energy involved in restoring the meter as waste and disregard it in computing the measurement energy cost. We note that unlike Landauer, in our model we use a non-thermal meter state whose preparation may require some energy investment \cite{footnote3}. To estimate this energetic cost: (a) Return the particle to its ground state by contacting it with a zero-temperature bath. (b) Adiabatically confine the particle in a harmonic potential of frequency $\Omega=\hbar B/2$ for which the wavepacket (meter state in Eq.\ \eqref{in1}) is the ground state. This step incurs an investment equal to the zero-point energy $\hbar \Omega/2$. (c) Suddenly remove the harmonic confining potential, a step that provides no energy gain. This set the wavepacket preparation energy cost as $W_{prep}=\hbar\Omega /2 = \hbar^2 B/4$ and the total energy for the measurement as 
\begin{align}
\label{invest2}
    W_{in}(t_m)= W_{meas}(t_m) + W_{prep}.
\end{align}

To show the dependence of these quantities on the measurement time $t_m$ it is convenient to express time in terms of a timescale defined by the coupling constant and the parameters of the initial meter wavepacket Eq.\ \eqref{meterIn}. Henceforth, time is represented in terms of the reduced quantity $\bar{t}=t/\tau^*$ where $\tau^* = 2b \frac{\sqrt{\langle \delta \hat{p}^2( t=0) \rangle}}{|d \langle \hat{p}( t) \rangle / d t|_{t=0}} = \frac{\sqrt{\hbar^2 B}}{ g}$, and where $|d\langle \hat{p} ( t) \rangle/d t|_{ t=0}$ is the initial change rate of the meter momentum (see \cite{SM2}). 

\textit{Information gain and energetic cost- }
 Fig.\ \ref{fig9} (A) illustrates the conditional probability $P_{i=0;1}(t_m|p)$ to be in the excited state given the meter outcome $p$. Obviously, $P_{i=1,2}(t_m=0|p) = a, b$ is independent of $p$. For $t_m>0$, the evolution of these probabilities may be written as $a\to a-\delta$ and $b\to b+\delta$, where, if $g$ is chosen positive, $\delta>0$ if the meter outcome is negative ($p<0$), and $\delta <0$ when $p>0$, indicating a higher or lower likelihood that the 2SS is in the excited state, respectively. The information gain $I(t_m)$ (Eq.\ \eqref{know}) and measurement cost $W_{meas}(t)$ (Eq.\ \eqref{invest2} first term) are depicted in Fig.\ \ref{fig9} (B) as function of $t_m$ for different initial 2SS states defined by $b/a=\exp{[-\Delta E /k_BT_S]}$. Three observations are notable:

(i) The information gain is a monotonously increasing function of $t_m$ that approaches the entropy of the initial state, $-k_B(a\ln a +b \ln b)$, as $t_m\to \infty$. This stands in contrast to the model of Ref.\ \cite{bre2021} where, because of the discrete nature of the meter (another two state system), the dependence on $t_m$ reflects the intrinsic Rabi-oscillation in the system-meter dynamics.

(ii) The rate of information gain, given by the slope $dI(\bar{t}_m)/d\bar{t}_m$ of $I(\bar{t}_m)$ in Fig.\ \ref{fig9} (B), is maximal near $t_m=\tau^*$, a time determined by the width of the initial meter wavepacket and the system-meter coupling. 

(iii) As measurement time $t_m$ increases the information gain $I$ approaches its maximal value. However, in the model considered, the measurement energy cost $W_{meas}$ (first component in Eq.\ \eqref{invest2}) increases indefinitely, (see dotted lines in Fig.\ \ref{fig9} (B)) resulting in a decreasing trend of the information gain to energy ratio. As already stated, in realistic settings $t_m$ may be controlled by the range of system-meter interaction.

\textit{Efficiency and Power output-}
We define the IE's efficiency of work extraction  by \cite{footnote8} $\eta (t_m,p') =W_{out} (t_m,p')(Q_{in} (t_m,p')+W_{in}(t_m))^{-1}$, where $Q_{in}(t_m,p')$ is the average heat per cycle taken from the system thermal bath in order to return the 2SS back to thermal state following work (photon) extraction. This average heat is equal to the average work extracted per cycle $Q_{in}(t_m,p')\equiv W_{out}(t_m,p')$. Therefore
\begin{align}
\label{efficiency}
 \eta (t_m,p') =\frac{1}{1+W_{in}(t_m)/W_{out} (t_m,p')}. 
\end{align}
It is important to note that our idealized model disregards other physical processes in which part of this energy input might be lost, such as non-radiative decay of the 2SS. Eq.\ \eqref{efficiency} should therefore be regarded as an upper bound to the efficiency of a realistic engine. 
Also, the average power output of the engine per measurement cycle, is obtained from
\begin{align}
\label{power}
\Pi(t_m,p') &=\frac{ W_{out}(t_m,p') }{t_m}.
\end{align}

\begin{figure}[h!!!!!]
\centering
\includegraphics[width=\linewidth]{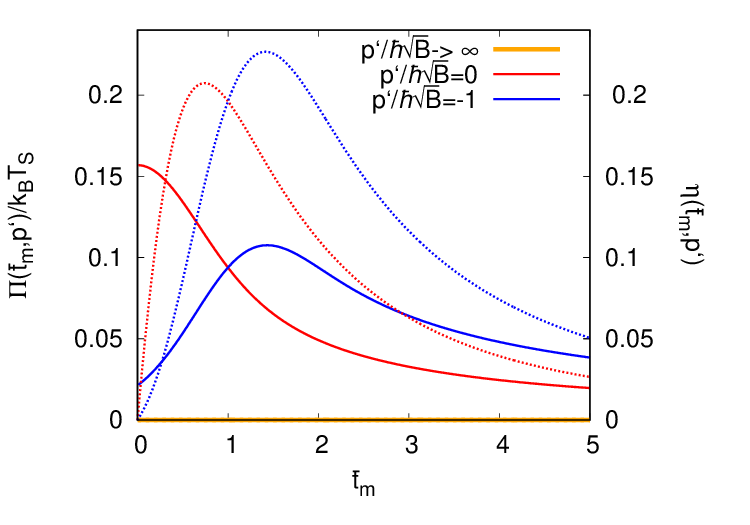}
\caption{\label{fig3} Output power $ \Pi(\bar{t}_m,p')$ (solid lines, left axis) and engine's efficiency $\eta (\bar{t}_m,p') $ (dotted lines, right axis) shown as functions of system-meter interaction time $\bar{t}_m$ for different values of threshold parameter $p'$ to attempt photon extraction by stimulated emission. Parameters are the same as those used in Fig.\ \ref{fig9} (A).}
\end{figure}

Fig.\ \ref{fig3} shows these efficiency $ \eta (t_m,p') $ and power output $\Pi(t_m,p')$, against the measurement time $t_m$ for different bounds $p'$ for triggering photon extraction attempt. Several observations follow:

(i) Choosing $p'\to \infty$, that is disregarding the measurement outcome in proceeding with photon extraction attempts, leads to vanishing efficiency and power, that is zero gain in the IE operation. Conversely, for $p'\to -\infty$, the 2SS is determined to be in its excited state with probability approaching $1$, so photon extraction attempt is assured. However, in this limit $t_m/t_{eff}$ in Eq.\ \eqref{gain2} vanishes, therefore $W_{out}(t_m,p'\to - \infty)=0$, implying $\eta(t_m,p'\to -\infty)=0$.

(ii) For a given finite $p'$, the IE efficiency increases as $t_m$ increases from zero (see Fig.\ \ref{fig3}), indicating a finite measurement time is needed for the IE operation. 

(iii) For the IE model considered, in which the system-meter interactions remains constant until cutoff at time $t_m$, the saturation in the time of information gain (see Fig.\ \ref{fig9}(B)) and increasing energy cost imply that the efficiency vanishes at $t_m\to \infty$. Consequently, $\eta(t_m,p')$ exhibits a peak at some intermediate measurement time. 

(iv) In the limit $\bar{t}_m\to 0$ the power output is given by $ \Pi (\bar{t}_m\to 0,p') =ab\Delta E\sqrt{\frac{2}{\pi}}e^{\frac{-2p'^2}{\hbar^2 B}}$ \cite{SM3}. For $p'\geq 0$, this is the maximal value of $\Pi$, which is a monotonously decreasing function of $t_,$ as exemplified by $p'=0$ in Fig.\ \eqref{fig3}.

(v) For $p'<0$, the average power output goes through a maximum as function of $t_m$. Its initial increase with $t_m$ reflects again the fact that the that the system and the meter has to interact for enough time to affect a useful work output. The decrease at long time results from the fact that the IE is extracting at most energy $\Delta E$ per cycle whose duration increases as $t_m \to \infty$ while information gain saturates. 
 
\begin{figure}
\centering
\includegraphics[width=\linewidth]{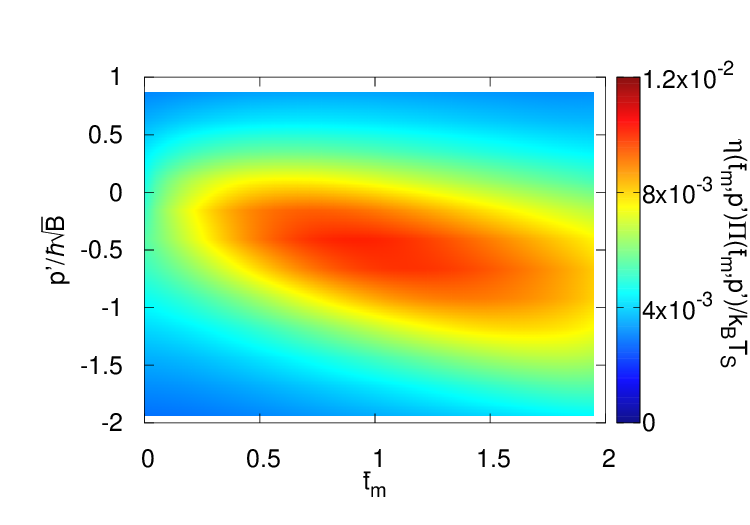}
\caption{\label{figIJ} Product of engine power and efficiency $ \eta (\bar{t}_m,p')  \Pi(\bar{t}_m,p') $ as color plot in dependence of upper bound of $p'$ where we send a photon in for potentially stimulating a photon emission and various system-meter interaction time $\bar{t}_m$. Parameters are the same as those used in Fig.\ \ref{fig9} (A).}
\end{figure}

As performance quantifiers, the efficiency $\eta$ and power $\Pi$ provide complementary views of machine operations. Their product, $ \eta (\bar{t}_m,p')  \Pi (\bar{t}_m,p') $, my be used as a balanced quantifier. The heat map in Fig.\ \ref{figIJ} displays this product against the measurement time $\bar{t}_m$ and the threshold parameter $p'$. Importantly, a finite measurement time is required for optimal operation. Note that both the efficiency and power trend to zero as $t_m\to \infty$ because of the runaway energy cost and saturation in work extraction and information gain in this limit.  

Finally, noting that in general the IE operates between two temperatures, it is interesting to compare its performance to that of a standard heat engine. First, unlike a heat engine, part of the energy input, $W_{in}$ of Eq.\ \eqref{invest2}, needed for the IE operation is 'useful' work, rather than heat. Therefore, $W_{out}/W_{in}>1$, is a performance requirement which in our model translates into $\eta > 1/2$ (see Eq.\ \eqref{efficiency}). While examples in Fig. \ref{fig3} fall below this limit, this criterion can be satisfied by increasing $\Delta E$ and $T_S$ while keeping their ratio $\Delta E/k_BT_S$ constant (see Fig. (S1) in \cite{SM1}). Second, the IE efficiency might exceed the Carnot limit $\eta_C=1-T_M/T_S$ only if $T_M$ is larger than some threshold value. We leave detailed considerations of this issue to future work. 

\textit{In summary,} we have analyzed an information engine (IE) model which highlights the importance of considering measurement time and the energy cost associated with its operation. Although details of these IE characteristics depend on the process used for information acquisition, their determination is required for estimating standard performance quantifiers such as engine efficiency and power. Our findings suggest that there is an optimal time for information acquisition beyond which increasing energy cost lead to diminishing returns. Using the product of efficiency and power as a performance quantifier we are able to identify the regime of optimal engine performance. In comparison to standard heat engine we find that our 2SS IE model is advantageous at large system energy gap and high system temperature, provided that its efficiency exceeds $0.5$. 
Moving forward, examination of other IE models with different working and measurement protocols are needed to delve further into the complexities of the measurement process, including entropic costs at finite temperatures of the measurement channel. This study lays the groundwork for investigating energy and entropy costs in realistic information-based engines and processes, particularly in quantum optic set-ups as proposed in previous research studies (e.g., \cite{cot2017}).

\begin{acknowledgments}
The research of A.N. is supported by the Air Force Office of Scientific Research
under award number FA9550-23-1-0368 and the University of Pennsylvania.
\end{acknowledgments}

\bibliography{MS}

\begin{thebibliography}{40}
\bibitem [{foo({\natexlab{c}})}]{footnote13}%
  \BibitemOpen
  \href@noop {} {\emph {\bibinfo {title} {This assumpltion has to be applied
  with caution: In thermal equilibrium no net gain can be achieved, implying
  that only the excess probability following measurement (relative to the
  thermal one) that the molecule is in the excited state must be associated
  with this net gain.}}}\BibitemShut {Stop}%
\bibitem [{foo({\natexlab{d}})}]{footnote14}%
  \BibitemOpen
  \href@noop {} {\emph {\bibinfo {title} {In making this identification we
  assume that the time needed for the photon extraction itself as well as the
  relaxation time associated with the restoration step discussed below can be
  disregarded. Otherwise these times should be added to $\Delta
  t$.}}}\BibitemShut {Stop}%
\end{thebibliography}


\end{document}